\documentclass[preprint2,openany]{aastex631}

\shorttitle{$\gamma$ rays from FR0 Radio Galaxies}
\shortauthors{Khatiya et al.}

\graphicspath{{./}{figures/}}
\usepackage{graphicx}
\usepackage{tikz}
\usepackage{float}
\usepackage{amsmath}
\usepackage{subfigure}
\usepackage{xcolor}
\usepackage{soul}
\usepackage{mathtools}
\usepackage{environ}
\NewEnviron{myequation}{%
\begin{equation}
\scalebox{1.1}{$\BODY$}
\end{equation}}
\newtagform{smalleqnum}[\small]{\small (}{)}
\defcitealias{Paliya:2021sbv}{P21}

\setstcolor{red}
\begin{document}

\title{Characterizing the $\gamma$-ray Emission from FR0 Radio Galaxies}

\author[0009-0002-2068-3411]{Nikita S. Khatiya}
\affiliation{Department of Physics and Astronomy, Clemson University, Clemson, SC 29634, USA}

\author[0000-0003-1046-1647]{Margot Boughelilba}
\affiliation{Universität Innsbruck, Institut für Astro- und Teilchenphysik, 6020 Innsbruck, Austria}

\author[0000-0002-6774-3111]{Christopher M. Karwin}
\affiliation{Department of Physics and Astronomy, Clemson University, Clemson, SC 29634, USA}
\affiliation{NASA Postdoctoral Program Fellow, NASA Goddard Space Flight Center, Greenbelt, MD, 20771, USA}

\author[0000-0002-8436-1254]{Alex McDaniel}
\affiliation{Department of Physics and Astronomy, Clemson University, Clemson, SC 29634, USA}

\author[0000-0002-7791-3671]{Xiurui Zhao}
\affiliation{Center for Astrophysics $\vert$ Harvard $\&$ Smithsonian, 60 Garden Street, Cambridge, MA 02138, USA}

\author[0000-0002-6584-1703]{Marco Ajello}
\affiliation{Department of Physics and Astronomy, Clemson University, Clemson, SC 29634, USA}

\author[0000-0001-8604-7077]{Anita Reimer}
\affiliation{Universität Innsbruck, Institut für Astro- und Teilchenphysik, 6020 Innsbruck, Austria}

\author[0000-0002-8028-0991]{Dieter H. Hartmann}
\affiliation{Department of Physics and Astronomy, Clemson University, Clemson, SC 29634, USA}

\begin{abstract}
FR0 galaxies constitute the most abundant jet population in the local Universe. With their compact jet structure, they are broadband photon emitters and have been proposed as multi-messenger sources. 
Recently, these sources have been detected for the first time in $\gamma$ rays. Using a revised FR0 catalog, we confirm that the FR0 population as a whole are $\gamma$-ray emitters, and we also identify two significant sources. For the first time, we find a correlation between the 5 GHz core radio luminosity and $\gamma$-ray luminosity in the 1 - 800 GeV band, having a 4.5$\sigma$ statistical significance. This is clear evidence that the jet emission mechanism is similar in nature for FR0s and the well-studied canonical FR (FRI and FRII) radio galaxies. Furthermore, we perform broadband SED modeling for the significantly detected sources as well as the subthreshold source population using a one-zone SSC model. Within the maximum jet power budget, our modeling shows that the detected gamma rays from the jet can be explained as inverse Compton photons. To explain the multi-wavelength observations for these galaxies, the modeling results stipulate a low bulk Lorentz factor and a jet composition far from equipartition, with the particle energy density dominating over the magnetic field energy density. 
\end{abstract}


\section{Introduction} \label{sec:intro}
Active galactic nuclei (AGNs) host accreting super-massive black holes (SMBHs) at their centers. The central engine can produce relativistic outflows, typically beamed into narrow cones and extending to the Megaparsec scale \citep{2019ARA&A..57..467B}. Particles in these relativistic jets are accelerated to near the speed of light by shocks.
The particle-particle and particle-photon interactions in these environments lead to emission of broadband electromagnetic radiation, and plausibly particle multi-messenger signals (i.e.~cosmic neutrinos and ultra-high-energy cosmic rays -- UHECRs) \citep{Dermer:2008cy, 2018Sci...361..147I, 2021APh...12802564M, 2022ApJ...933L..43B, 2022Sci...378..538I}. AGN can generally be classified depending on the observer's viewing angle using the Unified AGN Model \citep{1995}. Considering jetted-AGNs in particular, the class of sources viewed at angles $\lesssim 10^\circ$ to the relativistic jet axis are known as blazars, and the class of sources viewed at greater angles is termed radio galaxies (RGs). 
RGs can be further classified based on their extended radio morphology using the Fanaroff-Riley (FR) classification scheme \citep{1974MNRAS.167P..31F}. This scheme divides  RGs into FR type I galaxies (FRI) and FR type II galaxies (FRII). The latter class shows prominent hot spots at the end of their powerful radio lobes (i.e.~edge-brightened sources), whereas the former class is weaker and more diffuse (i.e.~edge-darkened sources) \citep{2021APh...12802564M}. Both classes of RGs are well known to emit high-energy $\gamma$ rays \citep{2009ApJ...707...55A,  2010Sci...328..725A, 2011ApJ...733...66I, DiMauro:2013xta, 2019A&A...627A.148A, 2019ApJ...879...68S,
2020MNRAS.496..903H,
2020MNRAS.492.4666R}.

While bright extended sources like FRI and FRII galaxies have historically been the primary focus of radio observations, it has been known since the 1970s that there exists a widespread presence of compact radio sources at the centers of early-type galaxies (see \citet{2018A&A...609A...1B,2020} and references therein). Thanks to the advent of large-area high-sensitivity optical and radio surveys, a new class of radio galaxies has now emerged--the so-called FR0 galaxies \citep{Ghisellini:2011zf,2015, 2018A&A...609A...1B, 2019MNRAS.482.2294B, 2020}. 

FR0 hosts are mostly luminous red early-type galaxies, with large black-hole masses ($10^8 - 10^9 \ \mathrm{M_\odot}$), and are spectroscopically classified as low-excitation galaxies, similar to FRI galaxies. Indeed, FR0s share the same nuclear and host properties as FRIs. However, the primary distinguishing factor of FR0 galaxies is their lack of extended radio emission (i.e.~less than kiloparsec scales). FR0s are, therefore, characterized by their high core dominance evident from multiband analyses \citep{2015, 2018MNRAS.476.5535T} and have compact jet sizes extending $\leq$ 10 kpc. Moreover, FR0s represent the dominant population of radio sources in the local Universe (redshift of z $<$ 0.05), having a number density roughly 5 times higher than FRIs. Because of their ubiquity, it is imperative to study and characterize the broadband electromagnetic radiation from FR0 galaxies, including their high-energy $\gamma$-ray emission. 

Many aspects pertaining to the nature of FR0 galaxies remain unclear. Specifically, it is unclear why these compact sources lack extended jets, why they are so prevalent in the local Universe, and how they relate to the other FR-type galaxies. One key factor that may help to better understand some of these properties is the large-scale environment in which these galaxies reside. In \citet{2020}, it was found that FR0s live in regions of lower galaxy density with respect to FRIs, which is the first significant difference found between the two populations, aside from the lack of extended radio emission. Based on this result, the authors suggest a connection between the environment and jet power, driven by a common link with the black-hole spin. However, their analysis of the environment is only based on optical data, and a complementary study is still needed with X-ray observations. Additional uncertainties relating to FR0s include the nature of their central engine and source of radio emission (and whether it differs significantly from typical FRI and/or FRII galaxies), as well as their possible contribution to UHECRs, cosmic neutrinos, and the extragalactic $\gamma$-ray background \citep{Tavecchio:2017utw, 2019MNRAS.482.2294B,
2019ApJ...879...68S, 2021APh...12802564M}. 

High-energy $\gamma$ rays serve as a further probe of these open questions. 
The first claim of high-energy $\gamma$-ray emission from an FR0 galaxy was made in \citet{Grandi:2015eos}. Evidence was found for an association between the $\gamma$-ray source, 3FGL J1330.0$-$3818, and the FR0 galaxy, Tol 1326$-$379. This possible association was recently reevaluated in \citet{Wen-Jing:2021vjy}. The new analysis benefits from more statistics, improved instrument response (Pass-8), and an updated \emph{Fermi} point source catalog (4FGL). The position of Tol 1326$-$379 is found to be well outside the 95\% localization uncertainty of the 4FGL source, although still within the 68\% containment radius (reported to be 0$.6^\circ$ at 1 GeV) \citep{2009ApJ...697.1071A}. Moreover, we also note that Tol 1326$-$379 is not included in the sample of FR0 galaxies, denoted as FR0CAT \citep{2018A&A...609A...1B}. 
Most recently, \citet{Paliya:2021sbv} (hereafter \citetalias{Paliya:2021sbv}) reported the significant $\gamma$-ray detection of three FR0 galaxies, as well as a significant detection of the subthreshold sources, using a stacking technique. If confirmed, this would be the first bona fide $\gamma$-ray detection, opening a new window into the nature of FR0 galaxies.
In this work, we make a detailed study of the $\gamma$-ray properties of the FR0 population. We search for $\gamma$-ray emission from both individual sources and the subthreshold sample. This allows us to cross-check the results from \citetalias{Paliya:2021sbv}. Then, for the first time, we test for a correlation between the $\gamma$-ray luminosity and the core radio luminosity, in analogy to the well-known correlation that exists for FRI and FRII galaxies. 

Our analysis proceeds as follows. $\S$~\ref{sec:sample_selection} provides details for our sample selection. $\S$~\ref{sec: Data Analysis} covers the \emph{Fermi}-LAT 
data selection and data analysis. The main results are presented in $\S$~\ref{sec: Results}. Detailed modeling of the broadband spectral energy distributions (SEDs) is conducted in $\S$~\ref{sec: Model}. Finally, $\S$~\ref{sec: Discussion & Conclusion} gives our discussion and conclusion. Throughout this analysis we assume the following cosmological parameters: $\mathrm{H_0}$ = 71 $\mathrm{km\ s^{-1}\ Mpc^{-1}}$, $\Omega_m$ = 0.27, $\Omega_{\Lambda}$ = 0.73, similar to the one used by \citet{2020ApJ...894...88A}.

\section{Sample Selection} 
\label{sec:sample_selection}
We use the FR0 sample from FR0CAT. Note that the sample initially consisted of 108 sources, however, further analysis in \citet {2019MNRAS.482.2294B} indicated that four of them were spurious detections. Therefore, the final FR0CAT includes 104 sources, having z $\leq$ 0.05 with radio sizes $\leq$ 5 kpcs. Their radio luminosity lies in the range 10$^{38}$ $\leq$ $\nu$L$_{1.4}$ $\leq$ 10$^{40}$ $\mathrm{erg \ s^{-1}}$ and their black-hole (BH) masses lie between 10$^8$ \textless \ M$_{\textrm{BH}}$ \textless \ 10$^9$ M$_\odot$.

\section{Data Analysis}
\label{sec: Data Analysis}
\subsection{Data Selection} \label{Data}

We analyze $\gamma$-ray data collected by the \emph{Fermi}- Large Area Telescope (LAT) between 4 August 2008 and 25 December 2020 (12.4 years). The events have energies in the range 1-800 GeV. The spatial bin size is 0.08$^\circ$. To reduce contamination from the Earth’s limb, we use a maximum zenith angle of 105$^\circ$. We deﬁne a 10$^\circ$ $\times$ 10$^\circ$  region of interest (ROI) centered at the position of each FR0 source. We use the standard data ﬁlters: DATA$\_$QUAL $>$ 0 and LAT$\_$CONFIG==1. The analysis is performed using FermiPy (v0.19.0), which utilizes the underlying Fermitools (v1.2.23). We select photons corresponding to the P8R3\_SOURCE\_v2 class \citep{2013ApJ...774...76A}. In order to optimize the sensitivity, we implement a joint likelihood analysis with the four point-spread function (PSF) event types available in the Pass 8 dataset. The data are divided into quartiles corresponding to the quality of the reconstructed direction, from the lowest quality quartile (PSF0) to the best quality quartile (PSF3). Since we use a binned likelihood analysis, each sub-selection has its own binned likelihood instance that is combined in a global likelihood function for the ROI. This is easily implemented in FermiPy by specifying the components section in the conﬁguration ﬁle. Each PSF type also has its own corresponding isotropic spectrum, namely, P8R3\_SOURCE\_V2\_PSF\emph{i}\_v1 (\emph{i} = 0$-$3). The Galactic diffuse emission is modeled using the standard component (gll\_iem\_v07), and the point-source emission is modeled using the 4FGL catalog (gll\_psc\_v20) \citep{2020ApJS..247...33A}. In order to account for photon leakage from sources outside of the ROI due to the PSF of the detector, the model includes all 4FGL-DR2 sources within a 15$^\circ$ $\times$ 15$^\circ$ region. The energy dispersion correction (edisp\_bins = $-$1) is enabled for all sources except the isotropic component.

\subsection{Analysis} \label{Analysis}

\citetalias{Paliya:2021sbv} reported three out of the original 108 FR0 sources as being significantly detected by the LAT, the rest remaining below the detection sensitivity.  Therefore, we conduct a stacking analysis of the updated FR0 subthreshold sample. This technique has been successfully implemented in studies of upper limits on dark matter interactions \citep{PhysRevLett.107.241302}, detection of the extragalactic background light \citep{2018Sci...362.1031F}, extreme blazars \citep{Paliya_2019}, star-forming galaxies \citep{2020ApJ...894...88A}, fast black-hole winds \citep{Fermi-LAT:2021ibj}, and molecular outflows \citep{McDaniel:2023vsh}. 

The stacking pipeline is divided into three main steps. In the first step, the ROI of each source is optimized using a likelihood fit. The test statistic (TS) is calculated as
\begin{equation}
\label{TS_def}
    \mathrm{TS}=-2 \ \mathrm{ln}(L_0/L),
\end{equation}
where \emph{L$_0$} is the maximum likelihood when the source is absent in the model, and \emph{L} is the likelihood when the same source is present in the model. The normalization and index of the Galactic diffuse component as well as the normalization of the isotropic component are left free in the fit. The normalization of sources with TS $\geq 25 $ within 5$^\circ$ of the ROI is also kept free. Both the index and normalization for sources with TS $\geq 500$ \ within 7$^\circ$ of the ROI are kept free. The target FR0 sources are fit with a power law spectral model, and the normalization and index are left free to vary in the fit. We also find new sources within the ROI using the \emph{find\_sources()} method in the FermiPy package. This method creates a TS map and uses an iterative source-finding algorithm on this map to detect new sources with a threshold of TS$>$16. The threshold for separation between two sources is set to 0.5$^\circ$. At the end of the first step, the pipeline also identifies any significant sources ($\mathrm{TS_{threshold}} \geq 25 $), which we then remove from the sample and analyze individually. 

The second step involves the construction of 2D TS profiles where we scan through the total integrated photon flux and photon indices for each FR0 galaxy. The scanning range for the total integrated photon flux is from 10$^{-13}$ to 10$^{-9}$ $\mathrm{ph \ cm^{-2} \ s^{-1}} $ (for $1-800$ GeV range) with 40 logarithmically spaced bins, and the photon indices range from $-1$ to $-3.3$ with a spacing of 0.1. The power law spectral model used for the fit is given by
\begin{equation} \label{PL2}
    \frac{dN}{dE}=\frac{N (\Gamma +1) E^\Gamma}{E_\mathrm{max}^{\Gamma+1}-E_\mathrm{min}^{\Gamma+1}}
\ , \end{equation} where N is the total integrated photon flux and $\Gamma$ is the photon index. All the parameters of the “background” point sources are fixed. The normalization of the isotropic component is again set free in this step. Similarly, the index and prefactor of the Galactic diffuse component are also set free in the fit. This scan is done for all event classes, which are added to get a single TS profile for each source. In the final step, the TS profile of each source is added to generate the final stacked TS profile.

\section{Results} \label{sec: Results}

\subsection{Flux Stacking} \label{sec:Flux stacking}
%%%%%%%%%%%%%%%%%%%%%%%%%%%%%%%%%%%%%%%%%%%%%%
\begin{figure*}[ht]
    \centering
    \includegraphics[width=\columnwidth]{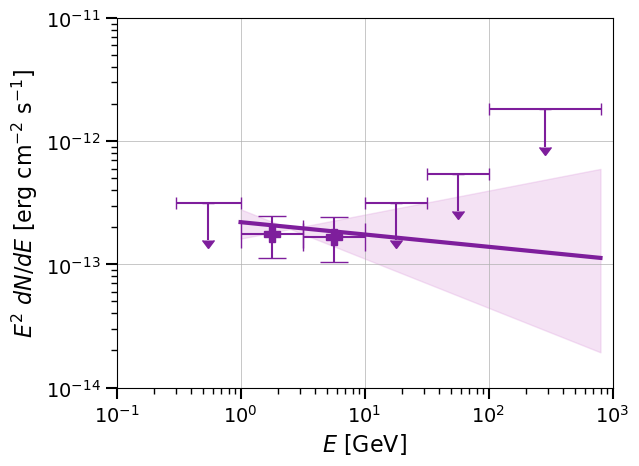}
    \includegraphics[width=\columnwidth]{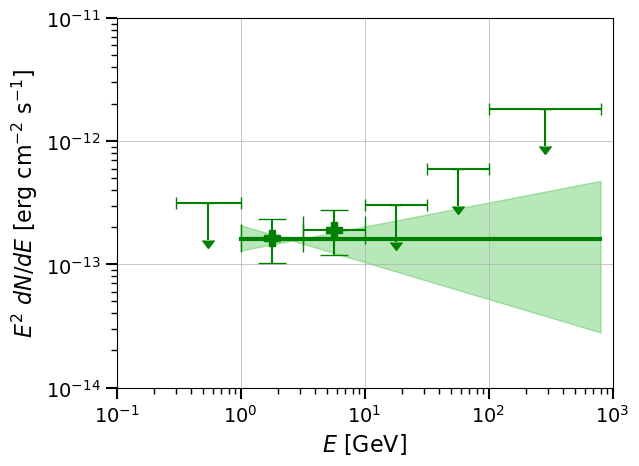}
    \caption{Left: Butterfly plot and SED for SDSS\_J153016.15+270551.0. The purple band shows the 1\,$\sigma$ uncertainty, and the solid purple line is the best-fit model. The purple data points show the corresponding SED calculated for different energy bins. Right: Butterfly plot and SED for SDSS\_J162846.13+252940.9. ULs in both panels are plotted at the 95 $\%$ C.L.}
 \label{SED_sig}
\end{figure*}
%%%%%%%%%%%%%%%%%%%%%%%%%%%%%%%%%%%%%%%%%%%%%%
From the updated sample of 104 FR0 sources, we detected two of them significantly in $\gamma$-rays, with spectral parameters and significance that are consistent with \citetalias{Paliya:2021sbv}'s work. One of the significant sources is SDSS\_J153016.15+270551.0 (also known as LEDA 55267), having a maximum TS of 35. Its $\gamma$-ray counterpart is listed in 4FGL-DR3 as 4FGL J1530.3+2709, with a separation of 3.35$^\prime$. This source is detected with a $\gamma$-ray photon flux of 1.25$_{-0.63}^{+0.74}$ $\times$ 10$^{-10}$ $\mathrm{ph \ cm^{-2} \ s^{-1}}$ (for $1-800$ GeV range) and spectral index of 2.1$_{-0.3}^{+0.4}$. The SED and butterfly plot for this source are shown in the left panel of Figure \ref{SED_sig}. The purple band is constructed by sampling the 2D TS profile, and the solid purple line corresponds to the best-fit values. The purple SED data points and $95 \%$ C.L. upper limits (ULs) are computed using the \emph{sed()} method in the FermiPy package. A TS threshold of 9 is used to plot ULs. For the low-energy bin, we use an energy range of 300 MeV to 1 GeV. The analysis is performed identically to the main analysis, with the exception that we set the maximum zenith angle to 100$^\circ$ to account for the larger PSF at low energy. 

The second significantly detected source is SDSS\_J162846.13+252940.9 (also known as LEDA 58287), having a maximum TS of 24. Note that we analyze this source separately since it is very close to our TS threshold, and it was significantly detected in \citetalias{Paliya:2021sbv}. The $\gamma$-ray photon flux is 1.0$_{-0.36}^{+0.58}$ $\times$ 10$^{-10}$ $\mathrm{ph \ cm^{-2} \ s^{-1}}$ and the spectral index is 2.0$_{-0.3}^{+0.4}$. We note that there is a nearby blazar from ROMA-BZCAT \citep{2008MmSAI..79..262M}, at a separation of 2.34$^\prime$. We, therefore, consider LEDA 58287 to be a candidate gamma-ray source, as also discussed in \citetalias{Paliya:2021sbv}.  The corresponding butterfly plot and SED are shown in the right panel of  Figure \ref{SED_sig}. 

We compared our results with the recent work in \citetalias{Paliya:2021sbv}. The author reports the detection of three significant sources, LEDA 55267, LEDA 58287, and LEDA 57137. For the first two sources, we find consistent spectral parameters and significance as \citetalias{Paliya:2021sbv}. However, the third source, LEDA 57137, has been removed from FR0CAT, since it was later found to have radio emission extending to roughly 267 kpc from the center of the galaxy, as discussed in \citet{2019MNRAS.482.2294B}. Therefore, we do not consider it in this analysis. 

\begin{figure}[t]
    \centering
    \includegraphics[width=\columnwidth]{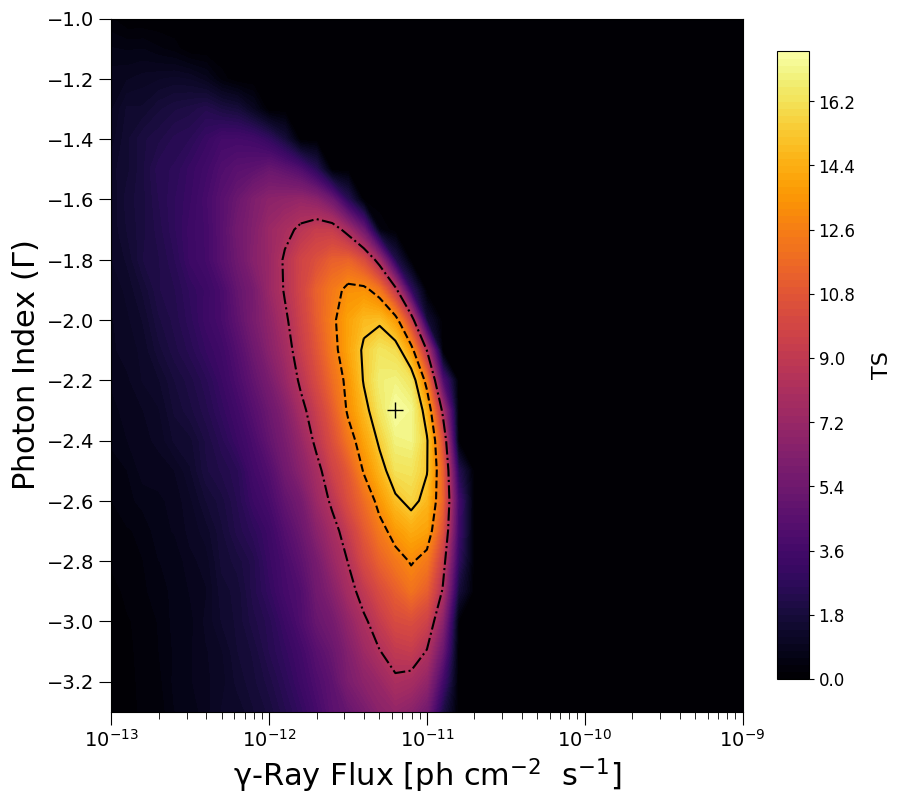}
    \caption{Stacked TS profile of the sub-threshold sources. The color scale indicates the TS value, and the + sign shows the maximum value. Significance contours are indicated in black for the 68\%, 95\%, and 99\% confidence levels.}
    \label{fig:1}
\end{figure}

\begin{figure}[t]
    \centering
    \includegraphics[width=\columnwidth]{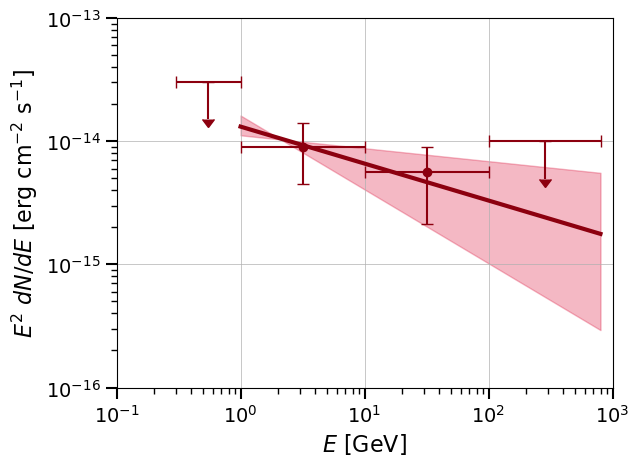}
    \caption{Butterfly plot for the sub-threshold sample. The red band shows the 1\,$\sigma$ uncertainty, and the solid red line is the best-fit model. The red data points show the corresponding SED calculated for four different energy bins. ULs are plotted at the 95 \% C.L.}
    \label{fig:2}
\end{figure}

The stacked TS profile for the subthreshold sample (102 sources) indicating the average best-fit parameters is shown in Figure \ref{fig:1}. They  are detected with a maximum TS of 17.53, giving a detection significance of 3.78\,$\sigma$, for two degrees of freedom. The best-fit $\gamma$-ray photon flux is 6.30$_{-2.33}^{+3.69}$ $\times$ 10$^{-12}$ $\mathrm{ph \ cm^{-2} \ s^{-1}}$ and the best-fit index is 2.3$_{-0.3}^{+0.3}$. The corresponding butterfly plot and SED are shown in Figure \ref{fig:2}. 

Our best-fit flux is a factor of 4.7 lower than that found in \citetalias{Paliya:2021sbv}. This is likely due to the different approaches used for the two studies. Our stacked profile is calculated using all sources in the sample, whereas \citetalias{Paliya:2021sbv} only selects sources with TS \textgreater \ 0, and then performs a background subtraction using blank sky regions. We note that our approach has been validated in previous studies \citep{Fermi-LAT:2021ibj}, i.e. using Monte Carlo simulations we are able to recover the input flux from the stacked profile. On the other hand, only selecting sources with TS \textgreater \ 0 would bias the results towards higher flux values. 
%%%%%%%%%%%%%%%%%%%%%%%%%%%%%%%%%%%%%%%%%%%%%%
%%%%%%%%%%%%%%%%%%%%%%%%%%%%%%%%%%%%%%%%%%%%%%

\subsection{The L$_{\gamma}$-L$_{radio}$ correlation} \label{sec:GR-correlation}

\begin{figure}[t]
    \centering
    \includegraphics[width=\columnwidth]{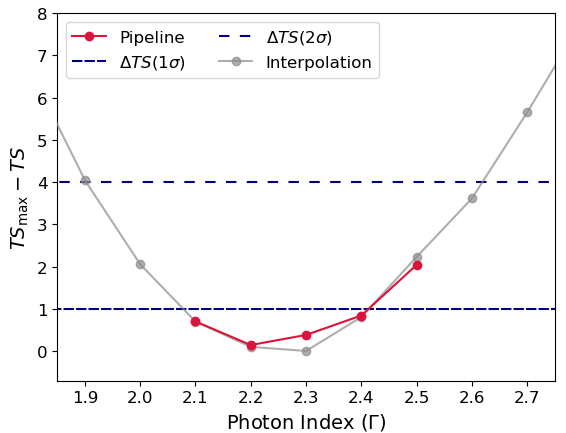}
    \caption{Index optimization for the $\alpha - \beta$ stacking. The grey curve shows the results from the interpolation technique. The red curve is generated from running the $\alpha-\beta$ stacking pipeline for indices near the minima of the grey curve.}
    \label{fig:3}
\end{figure}

\begin{figure}[t]
\begin{center}
 
\includegraphics[width=\columnwidth]{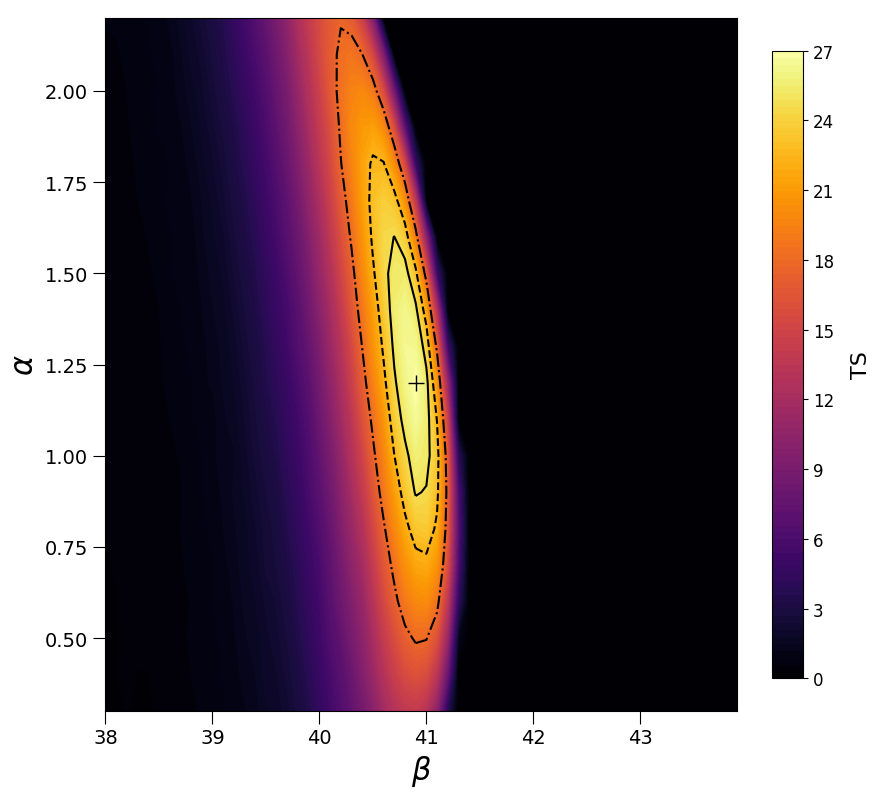}
 \caption{Stacked $\alpha-\beta$ profile for the sub-threshold sources. The color scale indicates the TS and the plus sign shows the best-fit $\alpha$ and $\beta$ values. Significance contours are indicated in black for the 68\%, 95\%, and 99\% confidence levels.}
\label{fig:4}
\end{center}
\end{figure}

For FRIs and FRIIs there is a known scaling relationship between the $\gamma$-ray luminosity and the 5 GHz core radio luminosity \citep{2011ApJ...733...66I, DiMauro:2013xta, 2019ApJ...879...68S}. We test if an analogous relationship exists for FR0 galaxies. We start by finding the correlation between the $\gamma$-ray luminosity and the 1.4 GHz total radio luminosity, which we then transform to the correlation with the 5 GHz core radio luminosity. Specifically, we test the relation
\begin{equation}
\label{correlation}
    \mathrm{log \ {L_\gamma}}=\beta + \alpha \ \mathrm{log}\left(\frac{\mathrm{\nu L_{1.4 GHz, total}}}{10^{39} \ \mathrm{erg \ s^{-1}}}\right), 
\end{equation} 
where $\mathrm{L_{\gamma}}$ is the $\gamma$-ray luminosity, $\mathrm{L_{1.4GHz, total}}$ is the luminosity density at a frequency ($\nu$) of 1.4 GHz (from FR0CAT), and $\alpha$ and $\beta$ are the scanned parameters in the fit. The normalization value of $10^{39} \mathrm{\ erg \ s^{-1}}$ is roughly equal to the mean radio luminosity of the sample. $\mathrm{L_{\gamma}}$ in 1 - 800 GeV energy band is evaluated using the following expression: 
\begin{equation}
\label{flux-lum}
    \mathrm{L_\gamma} = \mathrm{4  \pi}  d_l^2 (z) \ \frac{F_{\gamma}}{(1 + z)^{2 - \Gamma}}  
\end{equation}
where, $d_l^2 (z)$ is the luminosity distance at redshift z, $(1 + z)^{2 - \Gamma}$ is the K$-$correction factor and $F_{\gamma}$ is the energy flux given by $F_{\gamma} = \int_{E_{min}}^{E_{max}} E \frac{dN}{dE} \,dE $.  
%To properly take the distance of each source into account, we use their $\gamma$-ray luminosities and photon indices and perform the stacking in the $\alpha - \beta$ space.  
%%%%%%%%%%%%%%%%%%%%%%%%%%%%%%%%%%%%%%%%%%%%%%
%%%%%%%%%%%%%%%%%%%%%%%%%%%%%%%%%%%%%%%%%%%%%
The stacking in $\alpha - \beta$ space is performed the same way as described in Section \ref{Analysis}, except that instead of flux and index, we scan values of $\alpha$ and $\beta$. Specifically, we scan $\alpha$ values between $0.3 - 2.3$ and $\beta$ values between $38 - 44$, both having a bin size of 0.1. Since the conversion from flux to luminosity depends on the spectral index, we also scan through spectral indices. To save computational time, we first estimate the best-fit index by converting the flux-index plane (see Figure \ref{fig:1}) to the $\alpha-\beta$ plane using an interpolation method. This is done by scanning over multiple index values and the best-fit index is then determined from the likelihood profile. Finally, we run the stacking pipeline, using indices near the minimum of the converted likelihood profile, in order to robustly determine the best-fit index suitable for $\alpha-\beta$ stacking. Results for this are summarized in Figure \ref{fig:3}. From the $\alpha-\beta$ pipeline, we find a best-fit index of 2.2. This is consistent within 1\,$\sigma$ with the results from the flux-index plane. We note that the interpolation method prefers a slightly higher index of 2.3; however, we use the pipeline value since it is more robust. Moreover, the difference is very minor and does not make a significant change to the final results. 
Figure \ref{fig:4} shows the results for the sub-threshold sample. The corresponding best-fit $\alpha$ value is 1.2$_{-0.4}^{+0.3}$ and the best-fit $\beta$ value is 40.9$_{-0.2}^{+0.2}$, with a maximum TS of 26.99 corresponding to a significance of 4.5\,$\sigma$, for three degrees of freedom.

%%%%%%%%%%%%%%%%%%%%%%%%%%%%%%%%%%%%%%%%%%%%%%
%%%%%%%%%%%%%%%%%%%%%%%%%%%%%%%%%%%%%%%%%%%%%%

\begin{figure*}[t]
\begin{center}
\includegraphics[scale = 0.4]{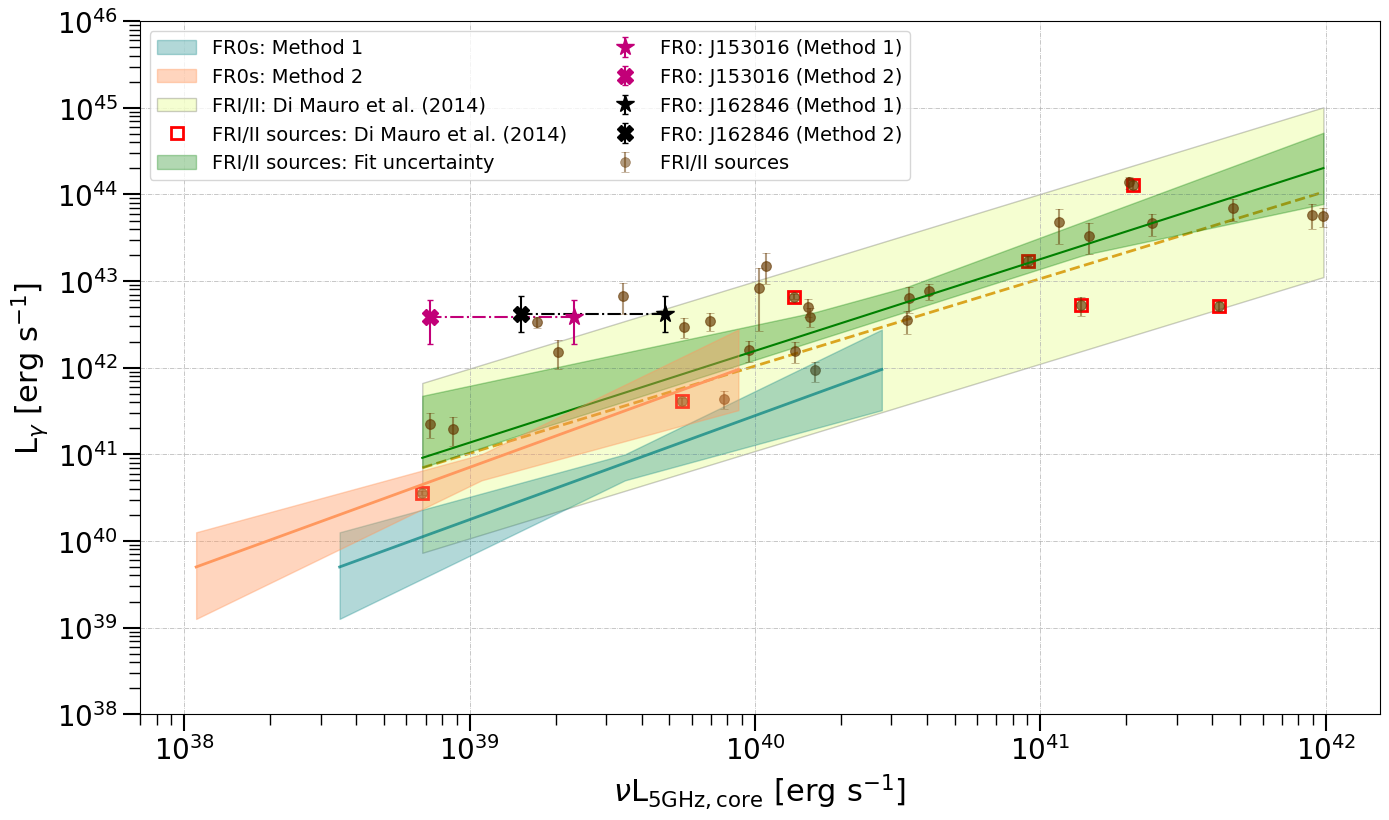} 
%width=0.9\textwidth
%\captionsetup{font=footnotesize}
\caption{Correlations between the $\gamma$-ray and 5 GHz core radio luminosities for FR0, FRI and FRII galaxies. The teal and orange butterfly plots correspond to the FR0 correlations from methods 1 and 2, respectively, as discussed in the text. The magenta and black markers are significant FR0 sources detected from the pipeline. The FRI/II sources are shown with brown-filled circles, and the green butterfly plot shows the uncertainty calculated from brown-filled circles using a Jackknife technique. The FRI/II sources used in \citet{DiMauro:2013xta} are indicated by open red squares, and the corresponding scaled L$_\gamma$ $-$ $\nu$L$\mathrm{_{5 GHz, core}}$ relation is shown with the yellow band. All errors except the green band are at 1$\sigma$ C.L.}
\label{lum correlation}
\end{center}
\end{figure*}

We use two methods for transforming L$_\mathrm{{1.4 GHz}, total}$ to L$_\mathrm{{5 GHz}, core}$. In the first method, we assume that FR0s behave as flat spectrum sources between $1.4 - 5$ GHz, as is the case for a majority of the FR0 population \citep{2019MNRAS.482.2294B, 2019A&A...631A.176C}. In addition, we assume that the total radio luminosity at 1.4 GHz all comes from the core, namely L$_\mathrm{{1.4 GHz}, total}$ = L$_\mathrm{{1.4 GHz}, core}$. This is motivated by the compact nature of FR0 sources as seen from high-resolution VLA observations, described in \citet{2019MNRAS.482.2294B}. This gives us the conversion of \begin{myequation}
\label{Method_1}
    \mathrm{L_{5 GHz, core}} = 3.57 \times \mathrm{L_{1.4 GHz, total}}
\end{myequation} The corresponding L$_\gamma$ $-$ L$\mathrm{_{5 GHz, core}}$ correlation for the subthreshold sources is shown in teal in Figure \ref{lum correlation}. 
%%%%%%%%%%%%%%%%%%%%%%%%%%%%%%%%%%%%%%%%%%%%%%
%%%%%%%%%%%%%%%%%%%%%%%%%%%%%%%%%%%%%%%%%%%%%%
%%%%%%%%%%%%%%%%%%%%%%%%%%%%%%%%%%%%%%%%%%%%%%
%%%%%%%%%%%%%%%%%%%%%%%%%%%%%%%%%%%%%%%%%%%%%%

To account for the fact that a portion of the total radio luminosity may come from outside the core region, we also use a second method where we adopt a core fraction from the literature. Specifically, we use the results from  \citet{Whittam_2017} for sources with L$_\mathrm{{1.4 GHz}, total}< 10^{25} \ \mathrm{W \ Hz^{-1}}$. From empirical simulations, the authors find a core fraction $\sim$ 0.3 at 1.4 GHz for flat spectrum sources. This gives us a conversion of 
\begin{myequation}
\label{Method_2}
    \mathrm{L_{5 GHz, core}} = 1.10 \times \mathrm{L_{1.4 GHz, total}}
\end{myequation}
The corresponding L$_\gamma$ $-$ L$\mathrm{_{5 GHz, core}}$ correlation is shown in orange in Figure \ref{lum correlation}.
The teal and orange butterfly plots indicate the two extremes where we expect the actual FR0 luminosity correlation to lie, given the uncertainty in the core radio luminosity at 5 GHz. In Figure \ref{lum correlation}, we also include the L$_\gamma$ and L$\mathrm{_{5 GHz, core}}$ measurements for the two significantly detected FR0 sources, using equations \eqref{Method_1} and \eqref{Method_2}. Results from \eqref{Method_1} are shown with magenta and black stars and those from \eqref{Method_2} are shown with crosses of the same colors. Method 1 and Method 2 results for these data points are connected by the dot-dashed lines, indicating the uncertainty in the L$_\mathrm{{1.4 GHz}, total}$ and L$\mathrm{_{5 GHz, core}}$ conversion from the two methods. 

In order to analyze the L$_\gamma$ $-$ L$\mathrm{_{5 GHz, core}}$ correlation for all three classes of FR radio galaxies, we further acquired a sample of FRI/II sources detected in the \textit{Fermi}-LAT 4FGL-DR2 catalog. In the catalog, 44 sources have an FRI/II classification. Here we use 32 of these sources, as 12/44 sources are excluded for various reasons, as stated below. Since our analysis focuses only on point sources, we exclude the extended sources Fornax A and the Centaurus A lobes. We exclude Centaurus B since it lies in the Galactic plane. The sources 3C 17, 3C 111, and PKS 2153-69 are excluded since they have negligible emissions in the 1-800 GeV energy band. Finally, PKS 0235+017, NGC 6328, PKS 2324-02, PKS 2327-215, and PKS 2338-295 are excluded since no L$\mathrm{_{5 GHz, core}}$ data are available.
We determine L$_\gamma$ for the remaining 32 FRI/II sources in the 1-800 GeV energy band by repeating the flux stacking pipeline and using the redshift independent distances from NASA/IPAC Extragalactic Database (NED). In cases where distances are not available, we used redshifts from the \textit{Fermi}-LAT 4LAC-DR2 catalog \citep{2020ApJ...892..105A, Lott:2020wno}. The redshifts and distances for the FRI/II sources are listed in Columns 3 and 4 of Table \ref{FR1/2 table}. The L$\mathrm{_{5 GHz, core}}$ values were found from various references shown in Column 6. The core luminosity was specifically chosen so as to avoid any contamination from the large extended emission present in FRI/II galaxies.

%\tikz[overlay,remember picture]
%\draw (current page.west) -- (current page.east);
The brown-filled circles in Figure \ref{lum correlation} show the individually detected $\gamma$-ray FRI/II sources. We determine the scaling relationship for these sources using a least-squares fit, expressed as
%\usetagform{smalleqnum}
\begin{equation} \label{FR12-correlation}
\mathrm{log L_\gamma} = \  \mathrm{A + B} \ \mathrm{log}\left(\frac{\nu \mathrm{L_{5GHz,core}}}{10^{40} \ \mathrm{erg \ s^{-1}}}\right),
\end{equation} 
where A = $42.19^{+0.33}_{-0.33}$ and B = $1.05^{+0.39}_{-0.39}$. The uncertainties in A and B are the Jackknife estimated standard errors. The Jackknife technique is selected as it is a robust technique to compute best fits and their uncertainties. In Figure \ref{lum correlation}, we plot the best-fit line, along with the green band that indicates the intrinsic scatter in the individually detected FRI/II sources emitting $\gamma$-rays. This is done, by computing for each value of L$\mathrm{_{5 GHz, core}}$ the maximum and minimum values allowed by the Jackknife technique for L$_\gamma$. 
%We first find a set of best-fit lines corresponding to each Jackknife estimate (i.e. slope and intercept of the correlation) using the Jackknife technique. The range of intercept and slope for the best-fit lines is: ($\mathrm{A_{min}, A_{max}}$) = ($42.09, 42.49$) and ($\mathrm{B_{min}, B_{max}}$) = ($0.70,1.22$). For each value of L$\mathrm{_{5 GHz, core}}$, we then select the maximum and minimum L$_\gamma$ from the set of best-fit lines that are created using the range of A and B values, to produce the green band.
We compare these results with the L$_\gamma$ $-$ L$\mathrm{_{5 GHz, core}}$ correlation for FRI/II sources found in \citet{DiMauro:2013xta}. These results were derived using 12 sources and calculating the $\gamma$-ray luminosity between 100 MeV - 100 GeV. We scaled the $\gamma$-ray luminosity of these sources to the 1 - 800 GeV energy band. The scaled relation is shown with a yellow best-fit line and the corresponding 1\,$\sigma$ uncertainty band (from the error in $\beta$). We measure a slightly higher slope due to a number of newly detected sources, but overall the results are in good agreement. 

\section{Broadband Data and Modeling} \label{sec: Model}

The $\gamma$-ray-to-5~GHz core radio correlation suggests the correspondingly emitting particle populations to be related and/or located in the same region. With FR0s belonging to the class of low-luminosity AGN with accretion likely in the advection dominated accretion flow (ADAF) mode, dense external radiation fields that could serve as targets for particle-photon interactions are not expected, since the photon density of any external radiation field is usually much lower than the internal jet radiation field. Hence, $\gamma$-ray production due to inverse Compton scattering of external target photon fields has not been added to our modeling. We will therefore probe the one-zone synchrotron-self Compton (SSC) emission model as the simplest of such kinds, to reproduce the broadband SED of the individually LAT-detected LEDA~55267 (Figure~\ref{LEDA55267_SSC}) and LEDA~58287 (Figure~\ref{LEDA58287_SSC}), as well as the average SED of the subthreshold sample (Figure~\ref{Subthreshold_SSC}). 

\subsection{Broadband SED}
To assemble the broadband SEDs we collected the available data for the 104 sources of the FR0CAT \citep{2019MNRAS.482.2294B} in the NASA Extragalactic Database\footnote{https://ned.ipac.caltech.edu/} (NED). In \citet{2018MNRAS.476.5535T} an additional sample of 19 FR0s was studied at X-rays, among which 11 were not in the FR0CAT. From these 11 sources, one is mentioned to be at the center of its cluster (namely J004150.47-0 in Abell85, \citealt{2018MNRAS.476.5535T}) and to avoid contamination from the cluster we do not include this source in the sample. We, therefore, added these 10 sources to our sample,  using the observational data from the NED and SSDC SED builder\footnote{https://tools.ssdc.asi.it/SED/}, to add additional data. Hence we built a sample of 114 FR0 sources. We excluded data points for which the NED comment stated uncorrected observations for known sources in the beam or internal error. Furthermore, we excluded data taken with IRAS, because targeted position ellipses did not overlap with our sources' positions. Due to the low spatial resolution of the UV images provided by Galaxy Evolution Explorer (Galex \citealt{2010AJ....139.1212S, 2012AAS...21934001S, Whittam_2017}), we follow \cite{2016MNRAS.457....2G} and treat these data points as upper limits. 

The X-ray data are used if taken with XMM-Newton, Neil Gehrels Swift Observatory, or Chandra telescopes, while observations from older instruments with a larger angular resolution are discarded. Most of the data points from Chandra come from the Chandra Point Source Catalog 2.0.1 \footnote{https://cxc.cfa.harvard.edu/csc/} \citep{2020AAS...23515405E} and we use the fluxes obtained in the modified source region (\textit{flux\_aper}) of the sources. In order to have the most complete data collection possible for the two individually gamma-ray detected sources, we built \emph{Swift}-XRT spectra using the online tool\footnote{https://www.swift.ac.uk/user\_objects/}. For LEDA 55267, we used \texttt{XSPEC} and a binning of 20 counts per bin to present the data.

LEDA 58287 (SDSS J162846.13+252940.9) was serendipitously covered by a {\it Chandra} observation (ID: 15356), which was taken in Dec.~07,~2012 using ACIS-S with an exposure of 2.96~ks. We reduced the {\it Chandra} data using the {\it Chandra}'s data analysis system (CIAO, \citet{2006SPIE.6270E..1VF}) version 4.14 and {\it Chandra} CALDB version 4.9.4. The level = 1 data were reprocessed using the CIAO \texttt{chandra\_repro} script. The source spectrum was extracted from a circular region centered at the source with a radius of 15\arcsec\ (corresponding to 90\% PSF); the background spectrum was extracted from a circular region near the source with a radius of 30\arcsec. The source spectrum, background spectrum, ARF, and RMF files were extracted using the CIAO \texttt{specextract} tool following standard procedures. 
Since the source is 11\arcmin\ off-axis with a very limited exposure time, it has a limited number of (nine net) photons. Thus, we grouped its spectrum with three photons per bin using the HEAsoft task {\tt grppha}. The source spectra were analyzed using the \texttt{XSPEC} software \citep{1996ASPC..101...17A} with a power law model using cstat \citep{1979ApJ...228..939C}. The best-fit photon index is $\Gamma$ = 2.1$^{+1.6}_{-1.1}$ and the measured 2--10~keV flux is 8.1$^{+14.7}_{-6.2}$~$\times$~10$^{-14}$~erg~s$^{-1}$~cm$^{-2}$ at the 1\,$\sigma$ C.L.

The $\gamma$-ray data points and its 1\,$\sigma$ uncertainty band, for significantly detected FR0 sources and subthreshold sources, are taken from $\S$~\ref{sec:Flux stacking}.

\subsection{SED Modeling}
\begin{figure}[t]
    \centering
    \includegraphics[width=\columnwidth]{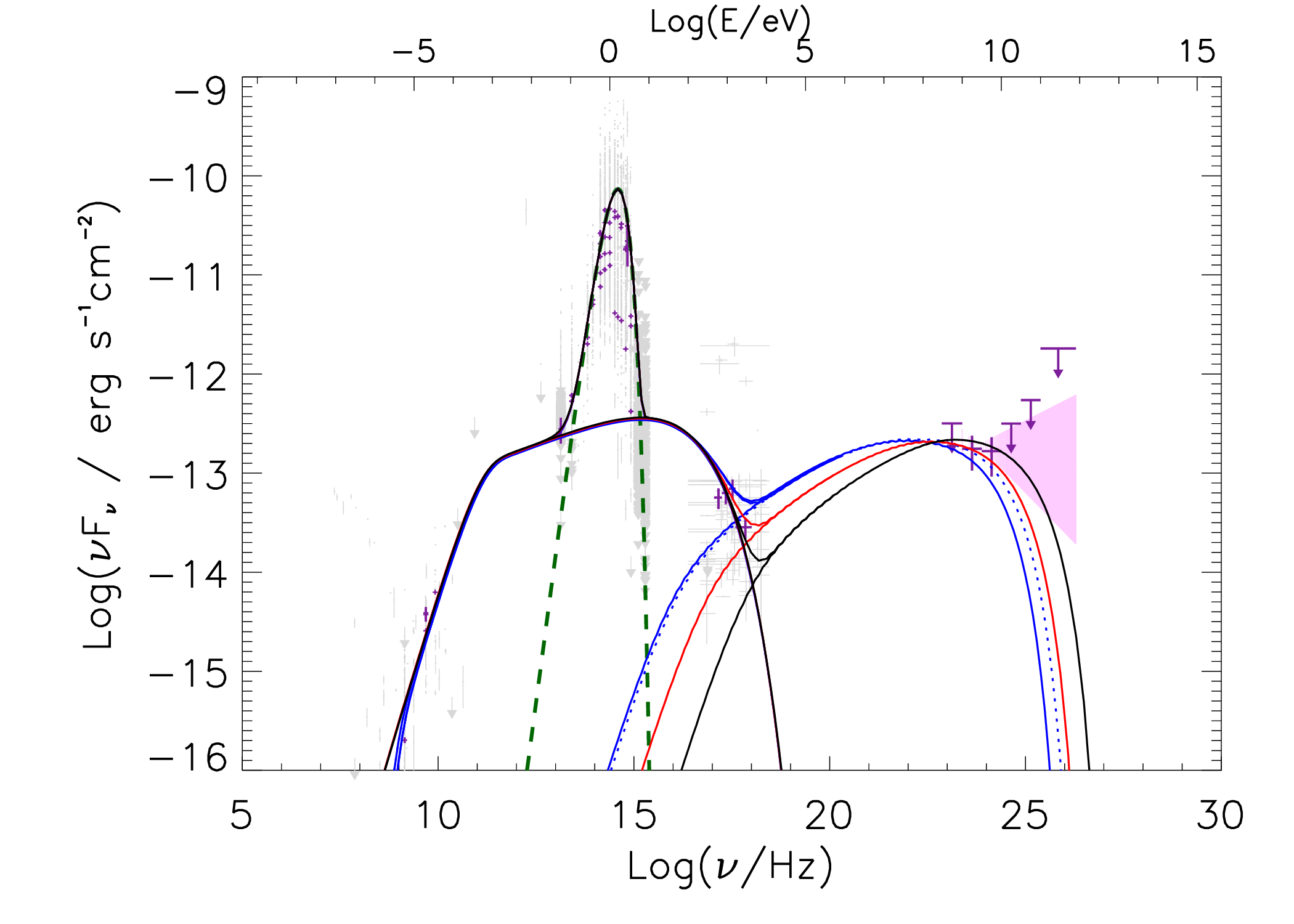}
    \caption{Broadband SED of LEDA~55267 (violet data points) on top of the SED data points of the entire FR0 sample (light grey data points), represented by one-zone steady-state SSC models with emitting exponential cutoff power-law particle spectra with index $p=2.7$, and that obey the jet power limit of $L_{\rm jet}\leq 10^{43.2}$$\mathrm{erg \ s^{-1}}$ (blue lines: $R=10^3r_g$, red lines: $R=10^4r_g$, black lines: $R=10^5r_g$, solid lines: $\Gamma=1.04$, dotted lines: $\Gamma=1.34$). Exponential cutoff particle energies reach $\sim (7\cdot 10^4 \ - \ 7\cdot 10^5) m_e c^2$ for the presented model SEDs. The host galaxy contribution is represented by a grey-body of temperature of $\sim 4800$K (dashed green line).}
 \label{LEDA55267_SSC}
\end{figure}

\begin{figure}[t]
    \centering
    \includegraphics[width=\columnwidth]{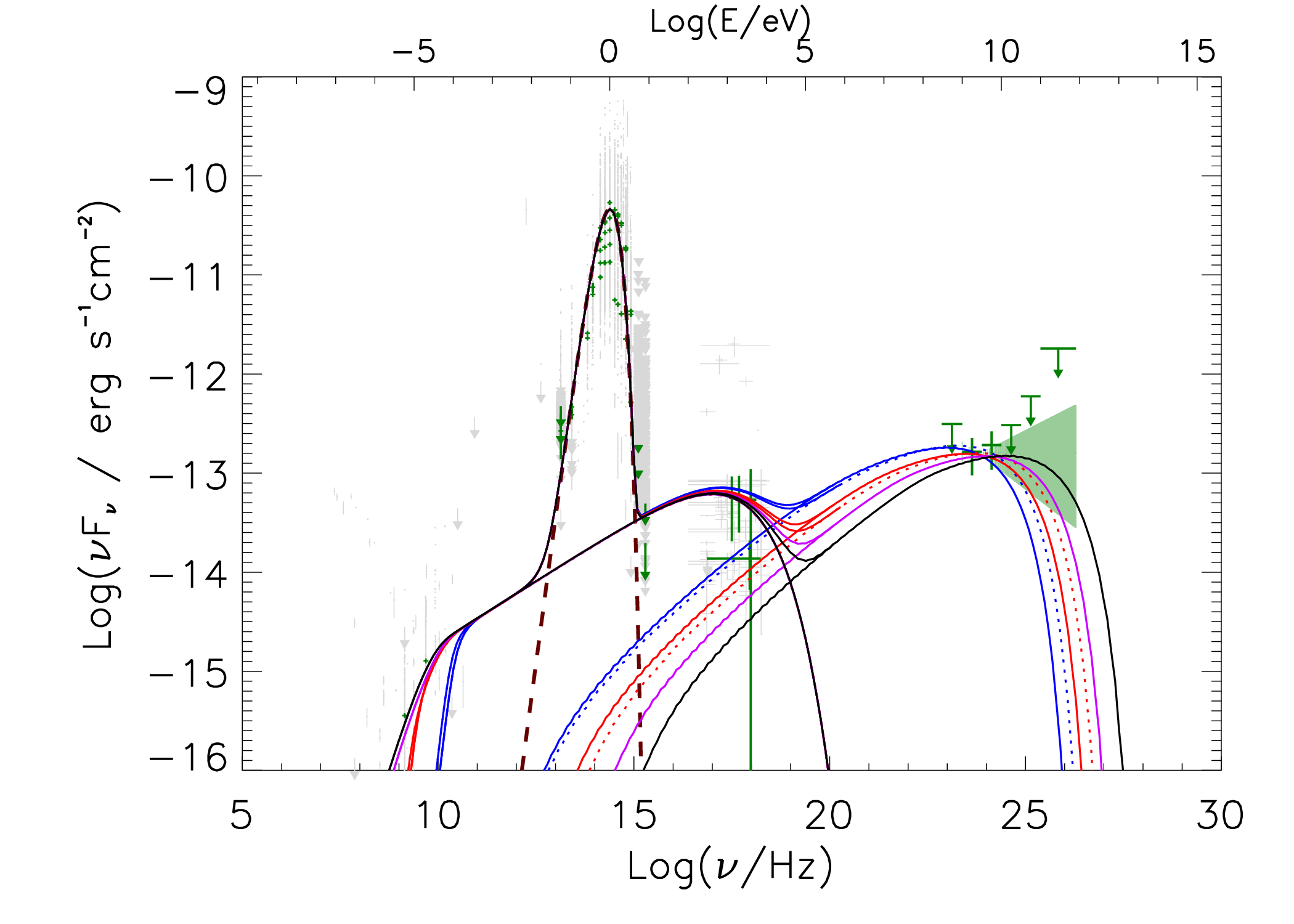}
    \caption{Broadband SED of LEDA~58287 (green data points) on top of the SED data points of the entire FR0 sample (light grey data points), represented by one-zone steady-state SSC models with emitting exponential cutoff power-law particle spectra with index $p=2.5$, and that obey the jet power limit of $L_{\rm jet}\leq 10^{43.6}$$\mathrm{erg \ s^{-1}}$ (blue lines: $R=10r_g$, red lines: $R=100r_g$, violet lines: $R=10^3r_g$, black lines: $R=10^4r_g$, solid lines: $\Gamma=1.04$, dotted lines: $\Gamma=1.34$). Exponential cutoff particle energies reach $\sim (1.3\cdot 10^5 \ - \ 6\cdot 10^6) m_e c^2$ for the presented model SEDs. The host galaxy contribution is represented by a grey-body of temperature of $\sim 4800$K (dashed brown line).}
 \label{LEDA58287_SSC}
\end{figure}

\begin{figure*}[ht]
    \centering
    \includegraphics[scale=0.16]{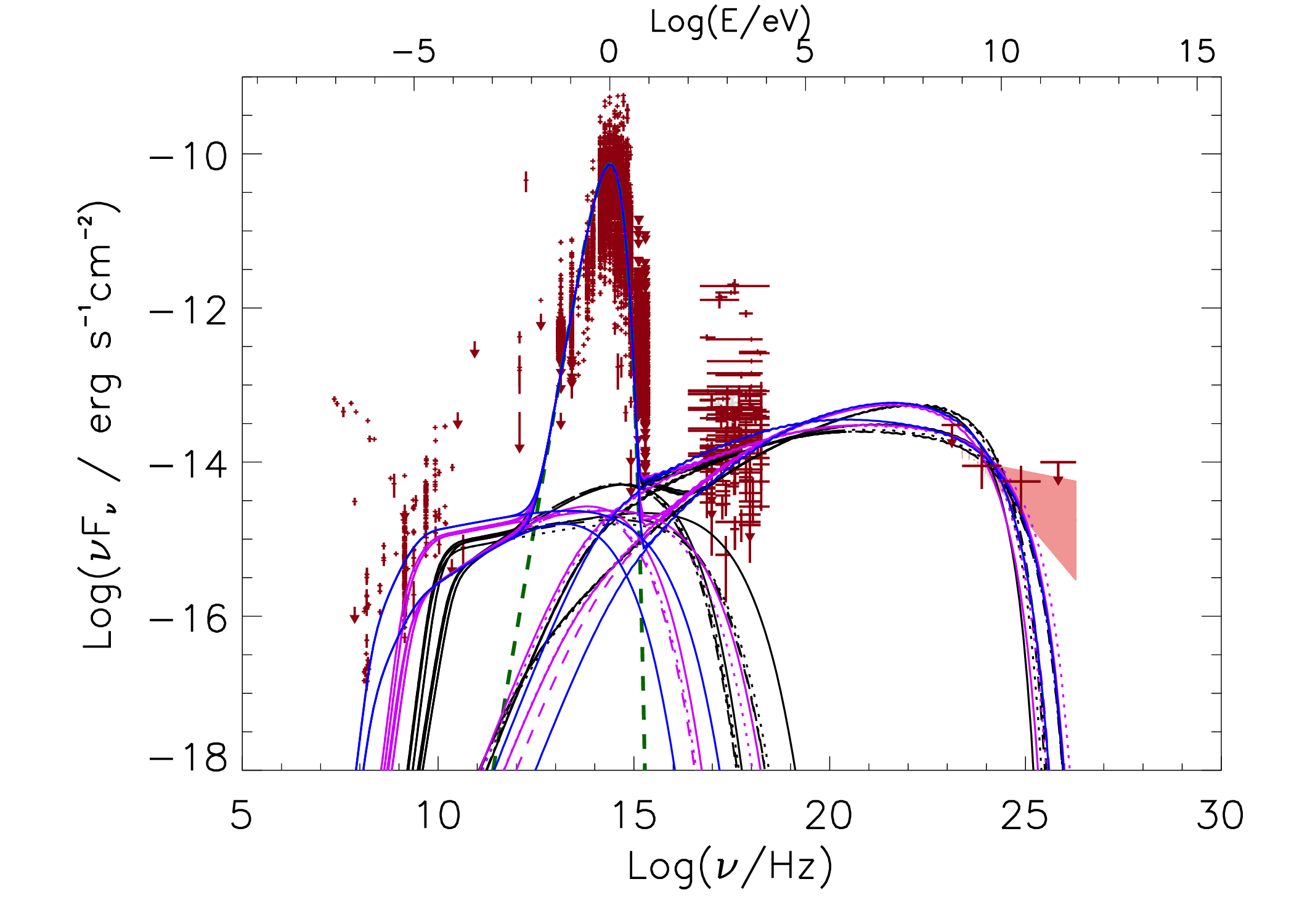}
    \caption{Broadband SED of the FR0 subthreshold sample (red data points) on top of the SED data points of  LEDA 55267 and LEDA 58287 (light grey data points), compared to one-zone steady state SSC models (see text for details) with emitting exponential cutoff power-law particle spectra with index $\alpha_e=2.3$ and $\alpha_e=2.8$,  that obey the jet power limit of $L_{\rm jet}\leq 10^{44}$$\mathrm{erg \ s^{-1}}$ (blue lines: $R=10^3r_g$, violet lines: $R=100r_g$, black lines: $R=10r_g$, solid lines: $\Gamma=1.04$, dotted lines: $\Gamma=1.34$, dashed lines: $\Gamma=2.34$ ). Exponential cutoff particle energies reach $\sim (10^4 - 10^5) m_e c^2$ for the presented model SEDs. The host galaxy contribution is represented by a grey-body of temperature of $\sim 4800$K (dashed green line). At X-ray energies, different flux data points at a given energy correspond to different sources. The $\sim 3$ orders of magnitude spread in X-ray flux corresponds roughly to the same spread in X-ray luminosity.}
 \label{Subthreshold_SSC}
\end{figure*}

The 1.4~GHz luminosity of LEDA~55267 and LEDA~58287 is used to estimate the corresponding average jet power of each source by means of the radio-to-jet power relation derived in \citet{2010ApJ...720.1066C}. Noting the 0.7~dex scatter of this relation we arrive at $L_{\rm{jet,max}}\approx 10^{43.2}$$\mathrm{erg \ s^{-1}}$ and $L_{\rm{jet,max}}\approx 10^{43.6}$$\mathrm{erg \ s^{-1}}$ for LEDA~55276 and LEDA~58287, respectively, for their maximum available jet power. The red elliptical host galaxies of the FR0s display prominently in their SEDs and are represented (dashed lines in Figs.~\ref{LEDA55267_SSC}-\ref{Subthreshold_SSC}) by a grey body with temperature $T\sim 4800$K as described in \citet{2021APh...12802564M}. The steady-state SSC model that is applied to the FR0 SEDs, considers a one-zone homogeneous, magnetized (with magnetic field strength $B$) spherical emission region (of radius $R$) that contains a relativistic electron population (with total emitting comoving electron energy density $u_e$) along with the same number of cold protons (with comoving energy density $u_p$). This region moves with relativistic speed $\beta c = c \sqrt{1-\Gamma^{-2}}$ ($\Gamma$ being the bulk Lorentz factor) along the jet axis that is itself misaligned to the line-of-sight. We chose a viewing angle of $20^o$, similar to the prototypical FRI M~87 (e.g., \citet{2019ApJ...875L...1E}). The emission model takes into account the full Klein-Nishina cross-section, exact expressions for the synchrotron and Compton emissivities, and synchrotron-self absorption. A power-law distribution (with index $p$) above a minimum electron Lorentz factor $\gamma_{\rm min}$ and with exponential cutoff $\propto \exp{(\gamma/\gamma_{\rm max})}$ for the emitting relativistic particle distribution is sufficient to describe the FR0 broadband SEDs adequately. 

Fig.~\ref{LEDA55267_SSC} shows a range of SSC models that represent the broadband data of LEDA~55267 within the given jet power constraint. Since most data lie in the radio-to-X-ray energy range, we first modeled these by the synchrotron emission of an electron distribution with power-law index $p=2.7$, scanning through a range of Doppler factors $D=1  -  3$ and region sizes $R=10^1  - 10^ 6$ gravitational radii $r_g$, and adjusted the field strength $B$ and energetics of the electron population to give the same synchrotron flux. Those corresponding SSC components that result in an overall satisfactory representation of all data are displayed in Fig.~\ref{LEDA55267_SSC}. We note that model SEDs with the same value for $(R B D^3)$ give the same SSC luminosity in the Thomson scattering regime (\citet{1998ApJ...509..608T}, \citet{2009ApJ...698.1761F}). For LEDA~55267, only mildly-relativistic jets ($\Gamma < 1.35$) with (sub-)pc emission region sizes $R\sim 10^3 - 10^5r_g$ are found to be compatible with the broadband data and within the jet power constraint. 

With a $\gamma_{\rm max} \sim 10^{4.8} - 10^{5.8}$ and $B\sim 10^{-2} - 10^0$G, the synchrotron spectrum extends here into the X-ray band. Lower values for $\gamma_{\rm max}$ that would potentially allow a SSC-interpretation of the X-ray data, cut off the SSC component below the $\gamma$-ray band, thereby leaving the LAT data unexplained in the given framework. The overall jet composition is found to be strongly particle dominated ($u_B/(u_e+u_p) \sim 0.07 - 0.3$). Similar conclusions were found from the successful SED modeling of LEDA~58287 using an emitting particle spectrum with index $p=2.5$, cutting off exponentially with $\gamma_{\rm max} \sim 10^{5.1} - 10^{6.8}$, contained in an emission region of size $R\sim 10^1 - 10^4 r_g$ that is moving along the jet axis with Lorentz factor $\Gamma<1.35$ (Fig.~\ref{LEDA58287_SSC}). For the rather weak field strengths required for these SSC models, the jet composition must be heavily particle dominated ($u_B/(u_e+u_p) \sim$ a few $10^{-4} - 10^{-3}$). 

The broadband data of the subthreshold sample sources (having a median redshift of $0.041$ and median black hole mass of $10^{8.3}M_\sun$) along with the results of an SSC modeling are summarized in Fig.~\ref{Subthreshold_SSC}. 
The spectral shape of the emitting particle distribution turns out to be mainly constrained by the LAT data, and the available jet power, estimated using again the radio-to-jet power relation of \citet{2010ApJ...720.1066C} and all available $\sim 20$~cm FR0 data (see also \citet{2019MNRAS.482.2294B}), to lie not beyond $10^{44}$ $\mathrm{erg \ s^{-1}}$.

Particle distributions harder than those with $p=2.3$ are found to not be compatible with the LAT spectral data points, while power-law distributions softer than with index $p=2.8$ quickly violate the jet power constraint. The normalization of the electron distribution (given the emission region size and Doppler-boosting) is chosen in order to not violate the flux upper limits in the optical/UV-band and at the same time meet the observed X-ray fluxes at least to some extent. While the X-ray flux for different FR0 sources may differ (see 
Fig.~\ref{Subthreshold_SSC}), temporal flux variability at these energies has never been 
recorded from any of the FR0s to our knowledge. The latter seems in line with interpreting the observed X-ray fluxes as SSC emission from rather low-energy electrons. Unlike the case of LEDA~55267 and LEDA~58287, the Compton dominance suggested by our modeling of the subthreshold sample lies below unity. Such SSC models typically require a low plasma magnetization, as is the case here ($B\sim 2\cdot 10^{-2} - 2\cdot 10^0 \ $G, and a strongly particle-dominated jet composition (here $u_B/(u_e+u_p) \sim 10^{-6} - 10^{-3}$). Overall the available jet power severely limits the allowed parameter space, e.g., only low bulk Lorentz factors $\Gamma<2.3$ and region sizes $R=10^1 - 10^3r_g$ are found in line with all constraints.

To summarize, the production mechanism of the gamma-ray radiation detected from FR0 galaxies is in line with inverse Compton scattering of jet synchrotron photons if the compact jet emission region is weakly magnetized and its total energy density is strongly particle-dominated. This emission site moves with only mildly relativistic speed along the misaligned jet axis, in agreement with the findings of \citet{2023A&A...672A.104G}.

\section{Discussion and Conclusion} \label{sec: Discussion & Conclusion}

Even though FR0s are fainter compared to the canonical FR-type galaxies, they are capable of emitting significantly at high energies up to the $\gamma$-ray regime. In this study, we confirm them as $\gamma$-ray emitters, with significant emission from two individual galaxies and an average $\gamma$-ray signal from the subthreshold sample. The two individually detected galaxies (SDSS J153016.15+270551.0 and SDSS J162846.13+252940.9) can be best modeled with photon fluxes of 1.25$_{-0.63}^{+0.74}$ $\times$ 10$^{-10}$ and 1.0$_{-0.36}^{+0.58}$ $\times$ 10$^{-10}$ $\mathrm{ph \ cm^{-2} \ s^{-1}}$ (for $1-800$ GeV range), spectral indices of 2.1$_{-0.3}^{+0.4}$ and 2.0$_{-0.3}^{+0.4}$ with a TS of 35 and 24, respectively, consistent with \citetalias{Paliya:2021sbv}. To characterize the $\gamma$-ray emission from the subthreshold sample, we utilized a stacking analysis technique,  which has been successfully implemented in a number of previous studies \citep{PhysRevLett.107.241302, 2018Sci...362.1031F, Paliya_2019, 2020ApJ...894...88A,Fermi-LAT:2021ibj,McDaniel:2023vsh}. We find that the subthreshold sample can be best described with an average photon flux of 6.30$_{-2.33}^{+3.69}$ $\times$ 10$^{-12}$ $\mathrm{ph \ cm^{-2} \ s^{-1}}$, average photon index of 2.3$_{-0.3}^{+0.3}$ with a detection significance of 3.78$\sigma$. This best-fit flux is a factor of 4.7 lower than \citetalias{Paliya:2021sbv}’s analysis due to a difference in both the analysis methods. Overall, our analysis shows that the two significant FR0 sources have a harder spectral index compared to the average population making them stronger $\gamma$-ray emitters.  

For the first time, we test the $\gamma$-ray and radio luminosity correlation for the FR0 source population, similar to the well-developed correlation for FRIs and FRIIs \citep{2011ApJ...733...66I, DiMauro:2013xta, 2016JCAP...08..019H, 2019ApJ...879...68S}. To investigate this, we implement $\alpha - \beta$ stacking on the subthreshold sample rather than the entire sample to avoid any biases due to the intrinsically bright FR0 galaxies. Stacking in the $\alpha - \beta$ space gives us an indication that the 5 GHz core radio and $\gamma$-ray band luminosities of FR0s are correlated with a significance of 4.5$\sigma$. Due to constraints on high-resolution 5 GHz core flux observations for all FR0 sources, the total luminosity at 1.4 GHz from the FR0CAT is initially used for the correlation and is then converted to 5 GHz core luminosity using two different assumptions that account for the uncertainty in the conversion. The true correlation should lie somewhere between the two extremes (see the teal and orange uncertainty bands in Figure \ref{lum correlation}). Furthermore, the luminosity correlation for FR0s is compared with the resolved population of $\gamma$-ray emitting FRIs and FRIIs. We refine the previous luminosity correlation for extended radio galaxies \citep{DiMauro:2013xta} using the updated 4FGL-DR2 catalog. The slope of the luminosity correlation for FR0s is $1.2_{-0.4}^{+0.3}$ and for FRI/FRIIs is $1.05_{-0.11}^{+0.11}$. 
The slope of this luminosity correlation for all three classes agrees well with each other within 1$\sigma$ errors. This indicates that the $\gamma$-ray luminosity scales with the core radio luminosity in a similar fashion for FR0s, FRIs, and FRIIs, which in turn closely relates to the jet emission mechanism. The correlation points towards the possibility that the underlying radiating particle populations are linked and/or experience the same environment within the jet. The two significant FR0 galaxies have a higher $\gamma$-ray luminosity, above the suggested luminosity correlation for the FR0 population. A smaller jet viewing angle ($\theta \sim 1/\Gamma$) may lead to higher $\mathrm{L}_\gamma$, however, we observe only two bright FR0s from a sample of 104 sources. The number of observed $\gamma$-ray bright FR0s is much lower than would be expected: ($2\pi\theta^2/4\pi$) $\sim12.5\%$ with $\Gamma \sim2$. This indicates that factors other than simple Doppler boosting could contribute to the high $\mathrm{L}_\gamma$ and need to be studied further. Deep and high-resolution radio imaging could put better constraints on the nature of the extended jets of these sources \citep{Lalakos:2023ean}. 

In recent studies, \citet{2022ApJ...931..138F} estimated $\sim 1-10\%$ contribution from all RG classes to the unresolved extragalactic gamma-ray background (EGB). Moreover, \citet{2019ApJ...879...68S} predict that core-dominated RGs, FR0s and FRIs, in total, contribute about $\sim 11 \pm 7 \%$ to the EGB. It is important to better constrain these estimates as FR0s are five times more abundant than the FRI and FRII classes. The derived luminosity correlation for FR0s will serve as a crucial ingredient, along with the previous comprehensive core luminosity function of RGs \citep{2018ApJS..239...33Y}, to get an updated and more precise estimate of the entire RG population contribution to the EGB.

Multi-band SED modeling reveals a wealth of information about the $\gamma$-ray emission mechanism and indirectly about the jet launching process in AGNs. Here, we consider a leptonic emission model with relativistic electrons close to the jet base which, in the presence of a magnetic field, undergoes synchrotron emission. The same population of relativistic charged particles scatter the low-energy jet synchrotron photons to high energies by the inverse Compton process, commonly called the SSC mechanism. To verify this presumption, we perform a one-zone SSC model fitting on the available multi-wavelength data for the two significantly detected FR0 sources and the subthreshold sample. In the absence of any observational constraints on the magnetic field strength at the FR0 jet base, we find this model in agreement with the current data. The GeV LAT data points, as well as the available jet power budget, notably constrain the allowed parameter values of this model. Our modeling suggests the emission region within a mildly relativistic jet to contain comparatively weak magnetic fields whose energy density is strongly sub-dominant as compared to the particle energy density. The synchrotron spectra of SDSS J153016.15+270551.0 and SDSS J162846.13+252940.9 extends up to the X-ray band, and the Compton dominance for these two galaxies lies around unity. For the subthreshold sample, the SSC model gives similar estimates for the bulk Lorentz factor and particle versus field energy content of the emitting regions of FR0 jets. The Compton dominance for the FR0 population as a whole lies below unity. Since the overall FR0 population can be explained by a SSC model, without the requirement of a hadronic component, it is still unclear if FR0s can be powerful cosmic ray accelerators and/or neutrino emitters. Further studies are needed to better constrain this possibility.    

While results from the luminosity correlation and multi-wavelength SED modeling for FR0s seem to be promising, there are still unanswered questions regarding the physics of their apparent link to the FRI and II galaxies, the structure and composition of FR0 jets, and the interaction with the circumgalactic environment to launch a scaled-down version of jets. Progress in this regard might be possible with state-of-the-art high-resolution and sensitive instruments such as the upcoming Square Kilometer Array (SKA). Very Long Baseline Interferometry (VLBI) images with greater details and observations at higher frequencies would be beneficial for disentangling the core and jet components in these galaxies. Constraints on the core magnetic field strength in FR0 jets could be obtained through polarization and Faraday rotation measurements with the Polarisation Sky Survey of the Universe's Magnetism (POSSUM), on the Australian SKA Pathfinder telescope (ASKAP). 

%\begin{acknowledgments}
\section*{Acknowledgements}
The \textit{Fermi}-LAT Collaboration acknowledges generous ongoing support
from a number of agencies and institutes that have supported both the
development and the operation of the LAT as well as scientific data analysis.
These include the National Aeronautics and Space Administration and the
Department of Energy in the United States, the Commissariat \`a l’Energie Atomique
and the Centre National de la Recherche Scientifique / Institut National de Physique
Nucl\`eaire et de Physique des Particules in France, the Agenzia Spaziale Italiana
and the Istituto Nazionale di Fisica Nucleare in Italy, the Ministry of Education,
Culture, Sports, Science and Technology (MEXT), High Energy Accelerator Research
Organization (KEK) and Japan Aerospace Exploration Agency (JAXA) in Japan, and
the K.~A.~Wallenberg Foundation, the Swedish Research Council, and the
Swedish National Space Board in Sweden.

Additional support for science analysis during the operations phase is gratefully
acknowledged by the Istituto Nazionale di Astrofisica in Italy and the Centre
National d’\`Etudes Spatiales in France. This work performed in part under the DOE
Contract DE-AC02-76SF00515.

This research has made use of the NASA/IPAC Extragalactic Database (NED),
which is operated by the Jet Propulsion Laboratory, California Institute of Technology,
under contract with the National Aeronautics and Space Administration. Clemson University is acknowledged for its generous allotment of compute time on the Palmetto Cluster.

AR and MB acknowledge financial support from the Austrian Science Fund (FWF) under grant agreement number I 4144-N27. MB has for this project received funding from the European Union’s Horizon 2020 research and innovation program under the Marie Sklodowska-Curie grant agreement No 847476. The views and opinions expressed herein do not necessarily reflect those of the European Commission. 

CMK's research was supported by an appointment to the NASA Postdoctoral Program at NASA Goddard Space Flight Center, administered by Oak Ridge Associated Universities under contract with NASA.

%\end{acknowledgments}
\software{This work benefited from the following softwares: \textsc{FermiPy} \citep{wood2017fermipy}, \textsc{Astropy} \citep{2022ApJ...935..167A},
\textsc{Scipy} \citep{2020SciPy-NMeth}, \textsc{Matplotlib} \citep{Hunter:2007}, \textsc{TOPCAT} (\url{http://www.starlink.ac.uk/topcat/}).}

\pagebreak
\bibliography{FRO_citations}{}
\bibliographystyle{aasjournal}
\nopagebreak[4]
\section{Appendix}

\startlongtable
\begin{deluxetable*}{lccc}
\tablecaption{List of FR0 radio galaxies \label{table:results1}}
\tablehead{ 
\colhead{Source Name}  & \colhead{Redshift} &  \colhead{Distance} & \colhead{$\mathrm{log \ L_r}$} \\ \colhead{} & \colhead{} & \colhead{[Mpc]} & \colhead{[erg/s]}}
\startdata
J010101.12-002444.4*  &  0.09674  &  316  &  - \\
J010852.48-003919.4  & 0.045 & 254 &  39.07  \\ 
J011204.61-001442.4  & 0.044 & 215 &  39.14 \\ 
J011515.78+001248.4  & 0.045 & 215 &  39.51 \\
J015127.10-083019.3  & 0.018 & 90.1 &  38.69 \\
J020835.81-083754.8  & 0.034 & 135 &  38.94 \\
J075354.98+130916.5  & 0.048 & 206 &  38.72 \\
J080624.94+172503.7*  & 0.10412 & 496 & - \\
J080716.58+145703.3  & 0.029 & 170 &  39.14 \\
J083158.49+562052.3  & 0.045 & 166 &  38.62 \\
J083511.98+051829.2  & 0.046 & 191 &  38.80 \\ 
J084102.73+595610.5  & 0.038 &  -  &   -  \\
J084701.88+100106.6  & 0.048 & 207 &  39.23 \\ 
J090652.79+412429.7  & 0.027 & 159 &  39.34 \\ 
J090734.91+325722.9  & 0.049 &  -  &  - \\
J090937.44+192808.2  & 0.028 & 129 &  39.28 \\
J091039.92+184147.6  & 0.028 & 103 &  38.95 \\ 
J091601.78+173523.3  & 0.029 &  -  &  - \\
J091754.25+133145.5  & 0.05 & 236 &  39.33 \\
J092405.30+141021.5*  & 0.1364 & -  &  - \\
J093003.56+341325.3  & 0.042 & 188 &  39.29 \\
J093346.08+100909.0  & 0.011 & 50.9 &  38.39 \\ 
J093938.62+385358.6  & 0.046 & 178 &  38.51 \\
J094319.15+361452.1  & 0.022 & 110 &  39.18 \\
J100549.83+003800.0  & 0.021 & 94.1 &  38.55 \\
J101329.65+075415.6  & 0.046 & 187 &  38.66 \\
J101806.67+000559.7  & 0.048 &  -  &  - \\
J102403.28+420629.8  & 0.044 & 199 &  38.60 \\
J102511.50+171519.9  & 0.045 &  -  &  - \\
J102544.22+102230.4  & 0.046 &  -  &  - \\
J103719.33+433515.3  & 0.025 & 93.6 &  39.29 \\
J103952.47+205049.3  & 0.046 &  -  &  - \\
J104028.37+091057.1  & 0.019 & 82.6 &  38.89 \\
J104403.68+435412.0  & 0.025 & 112 &  38.83 \\
J104811.90+045954.8  & 0.034 & 116 &  39.04 \\
J104852.92+480314.8  & 0.041 &  -  &  - \\
J105731.16+405646.1  & 0.025 & 108 &  38.94 \\
J111113.18+284147.0  & 0.029 &  -  &  - \\
J111622.70+291508.2  & 0.045 & 210 &  39.72 \\
J111700.10+323550.9  & 0.035 & 158 &  38.86 \\
J112029.23+040742.1  & 0.05 &  -  &  - \\
J112256.47+340641.3  & 0.043 & 194 &  39.02 \\
J112625.19+520503.5  & 0.048 & 235 &  38.92 \\
J112727.52+400409.4  & 0.035 & 185 &  38.89 \\
J113449.29+490439.4  & 0.033 & 150 &  39.09 \\
J113637.14+510008.5  & 0.05 & 206 &  38.81 \\
J114230.94-021505.3  & 0.047 & 183 &  38.69 \\ 
J114232.84+262919.9  & 0.03 & 142 &  39.15 \\
J114804.60+372638.0  & 0.042 &  -  &  - \\
J115531.39+545200.4  & 0.05 & 231 &  39.44 \\
J115954.66+302726.9*  & 0.10646 & 495  & - \\
J120551.46+203119.0  & 0.024 & 112 &  39.27 \\
J120607.81+400902.6  & 0.037 & 160 &  38.61 \\
J121329.27+504429.4  & 0.031 & 104 &  39.24 \\
J121951.65+282521.3  & 0.027 &  -  &  - \\
J122206.54+134455.9* & 0.0807 & - & - \\
J122421.31+600641.2  & 0.044 & 169 &  38.46 \\
J123011.85+470022.7  & 0.039 & 144 &  39.51 \\
J124318.73+033300.6  & 0.048 & 191 &  39.59 \\
J124633.75+115347.8  & 0.047 & 169 &  39.46 \\
J125027.42+001345.6  & 0.047 &  -  &  - \\
J125409.12-011527.1  & 0.047 & 191 &  38.67 \\
J125431.43+262040.6*  & 0.06929 & 294  & - \\
J130404.99+075428.4  & 0.046 & 165 &  38.68 \\
J130837.91+434415.1  & 0.036 &  -  &  - \\
J133042.51+323249.0  & 0.034 & 185 &  39.01 \\
J133455.94+134431.7  & 0.023 & 106 &  38.87 \\
J133621.18+031951.0  & 0.023 & 96.8 &  38.68 \\
J133737.49+155820.0  & 0.026 & 109 &  38.73 \\
J134159.72+294653.5  & 0.045 &  -  &  - \\
J135036.01+334217.3  & 0.014 & 54.2 &  38.70 \\
J135226.71+140528.5  & 0.023 & 101 &  38.64 \\
J135908.74+280121.3*  & 0.0646 & 257  & - \\
J140528.32+304602.0  & 0.025 & 119 &  38.24 \\
J141451.35+030751.2  & 0.025 & 77.6 &  38.43 \\
J141517.98-022641.0  & 0.047 & 207 &  39.13 \\
J142724.23+372817.0  & 0.032 & 143 &  38.85 \\
J143156.59+164615.4  & 0.048 & 175 &  38.65 \\
J143312.96+525747.3  & 0.047 & 225 &  39.12 \\
J143424.79+024756.2  & 0.028 & 144 &  38.40 \\
J143620.38+051951.5  & 0.029 & 132 &  38.74 \\
J144745.52+132032.2  & 0.044 & 205 &  38.67 \\
J145216.49+121711.5  & 0.031 & 136 &  38.39 \\
J145243.25+165413.4  & 0.046 & 198 &  39.06 \\
J145616.20+203120.6  & 0.045 & 192 &  39.20 \\
J150152.30+174228.2  & 0.047 & 215 &  39.16 \\
J150425.68+074929.7  & 0.049 &  -  &  - \\
J150601.89+084723.2  & 0.046 & 175 &  38.63 \\
J150636.57+092618.3  & 0.028 &  -  &  - \\
J150808.25+265457.6  & 0.033 & 188 &  39.08 \\
J152010.94+254319.3  & 0.034 & 155 &  38.87 \\
J152151.85+074231.7  & 0.044 & 218 &  38.97 \\
J153016.15+270551.0  & 0.033 & 172 &  38.82 \\
J153901.66+353046.0*  & 0.0779 & 374  & - \\
J154147.28+453321.7  & 0.037 &  -  &  - \\
J154426.93+470024.2  & 0.038 & 154 &  38.84 \\
J154451.23+433050.6  & 0.037 & 109 &  38.36 \\
J155951.61+255626.3  & 0.045 & 193 &  39.96 \\
J155953.99+444232.4  & 0.042 & 156 &  39.38 \\
J160426.51+174431.1  & 0.041 &  -  &  38.34 \\
J160523.84+143851.6  & 0.041 & 163 &  38.58 \\
J160641.83+084436.8  & 0.047 &  -  &  - \\
J161238.84+293836.9  & 0.032 & 154 &  39.04 \\
J161256.85+095201.5  & 0.017 &  -  &  - \\
J162146.06+254914.4  & 0.048 & 217 &  38.86 \\
J162846.13+252940.9  & 0.04 & 182 &  39.15 \\
J162944.98+404841.6  & 0.029 & 142 &  38.42 \\
J164925.86+360321.3  & 0.032 & 129 &  38.52 \\
J165830.05+252324.9  & 0.033 & 160 &  38.75 \\
J170358.49+241039.5  & 0.031 & 134 &  38.99 \\
J171522.97+572440.2  & 0.027 & 123 &  39.16 \\
J172215.41+304239.8  & 0.046 &  -  &  - \\
J235744.10-001029.9*  & 0.03106 & - &  - \\
Tol 1326-379*  & 0.02843 & -  &  - \\
\enddata
\tablecomments{Column 1: All source names start with a prefix \lq{SDSS}\rq \ (Additional 10 FR0 sources used only in the SED modeling are followed by $\ast$ symbol), Column 2: Redshift (from FR0Cat), Column 3: Redshift-independent distances (from NED) and Column 4: logarithm of 1.4 GHz luminosity calculated using Column 3.}
\end{deluxetable*}

\startlongtable
\begin{deluxetable*}{lccccc}
\tablecaption{ List of $\gamma$-ray detected FRI/II sources \label{FR1/2 table}}
\tablehead{ 
\colhead{Source Name}  & \colhead{Morphology} &  \colhead{Redshift} & \colhead{Distance}  & \colhead{log L$_\mathrm{{5GHz,core}}$} & \colhead{References} \\ \colhead{} & \colhead{} & \colhead{} & \colhead{[Mpc]} & \colhead{[erg/s]}}
\startdata
IC 1531 & FRI & 0.025 & 100 & 40.18 & \citep{1989MNRAS.236..737E}  \\ 
3C 17 & FRII & 0.220 & - & 42.53 & \citep{1993MNRAS.263.1023M} \\ 
NGC 315 & FRII & 0.016 & 68 & 40.23 & \citep{2001ApJ...552..508G} \\
TXS 0149+71 & FRI & 0.022 & - & 40.17 & \citep{2005PhDT.......126F} \\ 
PKS 0235+017 & FRI & 0.021 & 88.4 & - & \citep{2005PhDT.......126F} \\  
NGC 1218 & FRI & 0.029 & - & 40.93 & \citep{1993MNRAS.263.1023M} \\
B3 0309+411B & FRII & 0.136 & - & 41.90 & \citep{2005PhDT.......126F} \\
IC 310 & FRI & 0.019 & 74.4 & 39.81 & \citep{2019ApJ...879...68S} \\
NGC 1275 & FRI & 0.018 & 62.5 & 41.73 & \citep{2019ApJ...879...68S} \\
Fornax A & FRI & 0.005 & 16.1 & 37.98 & \citep{1993MNRAS.263.1023M} \\
4C +39.12 & FRI & 0.020 & - & 39.83 & \citep{2001ApJ...552..508G} \\
3C 111 & FRII & 0.048 & - & 41.48 & \citep{2019ApJ...879...68S} \\
3C 120 & FRI & 0.033 & - & 41.60 & \citep{1993MNRAS.263.1023M} \\
Pictor A & FRII & 0.035 & - & 41.11 & \citep{1993MNRAS.263.1023M} \\
PKS 0625-35 & FRI & 0.055 & - & 41.28 & \citep{1993MNRAS.263.1023M} \\
NGC 2329 & FRI & 0.019 & 70.2 & 39.44 & \citep{2005PhDT.......126F} \\
NGC 2484 & FRI & 0.043 & 171 & 40.58 & \citep{2001ApJ...552..508G} \\
NGC 2892 & FRI & 0.023 & - & 39.22 & \citep{Kharb_2004} \\
NGC 3078 & - & 0.008 & 36 & 38.95 & \citep{1994MNRAS.269..928S} \\
B2 1113+29 & FRI & 0.047 & - & 39.98 & \citep{Nilsson} \\
3C 264 & FRI & 0.022 & 113 & 40.00 & \citep{2001ApJ...552..508G} \\
NGC 3894 & FRI & 0.011 & 45.6 & 39.90 & \citep{1984ApJ...287...41W} \\
NGC 4261 & FRI & 0.007 & 32.4 & 38.81 & \citep{Nilsson} \\
M 87 & FRI & 0.004 & 15.2 & 39.89 & \citep{2001ApJ...552..508G} \\
PKS 1234-723 & FRI & 0.024 & - & 39.96 & \citep{2018ApJS..239...33Y} \\
TXS 1303+114 & FRI & 0.086 & 352 & 41.19 & \citep{2019ApJ...879...68S} \\
PKS 1304-215 & FRII & 0.126 & - & 41.22 & \citep{2019ApJ...879...68S} \\
Cen A & FRI & 0.0017 & 4.04 & 39.38 & \citep{1993MNRAS.263.1023M} \\
Cen B & FRI & 0.013 & - & 40.62 & \citep{2005ApJS..156...13M} \\
3C 303 & FRII & 0.141 & - & 41.58 & \citep{2001ApJ...552..508G} \\
B2 1447+27 & FRI & 0.030 & - & 39.81 & \citep{2018ApJS..239...33Y} \\
PKS 1514+00 & FRI & 0.052 & - & 40.57 & \citep{2019ApJ...879...68S} \\
TXS 1516+064 & FRII & 0.102 & 441 & 41.04 & \citep{2019ApJ...879...68S} \\
PKS B1518+045 & FRI & 0.052 & 213 & 40.54 & \citep{Fan2003} \\
NGC 6251 & FRI & 0.024 & 97.6 & 40.16 & \citep{2019ApJ...879...68S} \\
NGC 6358 & - & 0.014 & - & - & - \\
PKS 1839-48 & FRII & 0.112 & - & 41.32 & \citep{2019ApJ...879...68S} \\
PKS 2153-69 & FRII & 0.028 & 114 & 40.4 & \citep{2011ApJ...740...29K} \\
PKS 2225-308 & FRI & 0.056 & - & 39.9 & \citep{1989MNRAS.236..737E} \\
PKS 2300-18 & FRII & 0.129 & - & 41.86 & \citep{1984MNRAS.207...55H} \\
PKS 2324-02 & FRII & 0.188 & - & - & - \\
PKS 2327-215 & FRI & 0.031 & - & - & - \\
PKS 2338-295 & FRI & 0.052 & - & - & - \\
\enddata
\tablecomments{Column 1: Source name, Column 2: FRI/II morphology, Column 3: Redshift (from 4LAC Fermi catalog), Column 4: Redshift-independent Distances (from NED), Column 5: logarithm of 5GHz core luminosity and Column 6: References to Column 5}
\end{deluxetable*}
%\begin{longtable}[H]
%\begin{flushleft}
%\begin{longtable}[htbp]{c c c c c c}

%\end{longtable}
%\end{flushleft}

\end{document}